\documentclass{PoS}
\usepackage{amsmath,amsthm, amssymb, latexsym}
\usepackage{graphicx}
\usepackage{color}
\usepackage{booktabs}
\usepackage{url}

\title{Tuning of the strange quark mass with optimal reweighting}

\ShortTitle{Tuning of the strange quark mass with optimal reweighting}

\author{\speaker{Bj{\"o}rn Leder}, Jacob Finkenrath\\
        Department of Physics, Bergische Universit{\"a}t Wuppertal \\ Gaussstr.~20, 42119~Wuppertal, Germany\\
        E-mail: \email{leder@physik.uni-wuppertal.de}}

\abstract{Quark mass reweighting can be used to tune the mass of dynamical quarks.
The basic idea is to use gauge field ensembles generated at some bare mass
parameters to evaluate observables at different bare sea quark masses. This
involves the computation of so called reweighing factors which are given as 
ratios of fermion determinants. In the case of simulations including the strange 
quark, reweighting can be used to improve the approach towards physical quark 
masses. Optimal reweighting strategies combine a change of the strange quark 
mass with a change of the light quark masses in order to minimize the 
fluctuations of the reweighting factor. We present numerical test of such 
strategies for recent CLS2 simulations and a software package for mass 
reweighting based on openQCD.
\begin{flushright} WUP 15-01\end{flushright}
}

\FullConference{The 32nd International Symposium on Lattice Field Theory,\\
		23-28 June, 2014\\
		Columbia University New York, NY}

\DeclareMathOperator{\e}{e}

\newcommand{\ev}[1]{\left\langle #1 \right\rangle}

\newcommand{\beann}{ \begin{eqnarray*}}
\newcommand{\eeann}{ \end{eqnarray*}  }

\newcommand{\obs}{\mathcal{O}}
\newcommand{\rD}[1]{\mathrm{D}[#1]}
\newcommand{\rO}{\mathrm{O}}
\newcommand{\Dm}{\Delta m}
\newcommand{\f}{\mathrm{1f}}
\newcommand{\Neta}{N_{\eta}}

\newcommand{\nf}{{n_f}}
\newcommand{\set}[1]{\{#1\}}

\newcommand{\var}{\mathrm{var}}

\newcommand{\fm}{\,\mathrm{fm}}

\newcommand{\mb}{{\overline{m}}}

\newcommand{\Tr}{\mathrm{Tr}}
\newcommand{\ff}{\mathrm{2f}}

\newcommand{\mK}{m_{\rm K}}
\newcommand{\mpi}{m_\pi}
\newcommand{\mev}{\text{MeV}}
\newcommand{\tm}{\text{tm}}
\newcommand{\Wii}{W_\text{(ii)}}
\newcommand{\Wiii}{W_\text{(iii)}}
\newcommand{\Wiiv}{W_\text{(i+iv)}}

\begin{document}

\section{Why mass reweighting?}

Lattice QCD simulations proceed to generate ensembles at small lattice
spacings and light quark masses close to or at their physical values. This is possible
because of the algorithmic improvements of the last decade. Nevertheless, the cost
to generate \emph{independent} gauge configurations grows, particularly so because
of the growing autocorrelation times when the lattice spacing is lowered \cite{algo:csd}.
The problem is further aggravated by the inclusion
of a sea strange quark, which leads to an enlarged mass tuning problem
(see for example \cite{Bruno:2014jqa}).

In such a situation it might be more cost efficient to change the masses of a given
ensemble using reweighting instead of generating a new one. This is because although
reweighting is expensive it only has to be done
for independent gauge configurations and therefore its cost does not increase with 
the autocorrelation times. Also the naively expected linear scaling with the volume
might be much milder in reality \cite{Luscher:2008tw,lat14:jacob}.

Reweighting simply means changing the probability $P_{a}(U)$ (``weight'') of each gauge
configuration $U$ in the expectation value
\begin{equation} \label{e:exp-value}
   \ev{\obs}_{a} = \frac{1}{Z_{a}}\int\rD{U}\;P_{a}(U)\; \obs(U)\,, \quad
    P_{a}(U) = \e^{-S_g(\beta,U)}\,\prod_{i=1}^{\nf} \det(D(U)+m_i+i\gamma_5\mu_i)\,.
\end{equation}
Here we assume $\nf$ quark flavors with (possibly degenerate) bare masses $m_i$ and
twisted masses $\mu_i$, the gauge action $S_g$ only depends on $\beta$ and
 normalization $\int\rD{U}P(U)/Z \equiv 1$.
The subscript on the expectation value and the probability distribution specifies
the bare parameter set $a=\set{\beta,m_1,m_2,\ldots,m_\nf, \mu_1,\ldots,\mu_\nf}$.
Now an observable at bare parameter set
$b=\set{\beta',m_1',m_2',\ldots,m_\nf',\allowbreak \mu_1',\ldots,\mu_\nf'}$
is obtained from
\begin{equation}
   \ev{\obs}_{b} = \frac{\ev{\obs W}_a}{\ev{W}_a}\,,\quad W = \frac{P_b}{P_a}\,.
\end{equation}
If one restricts oneself to changes in the masses (hence $\beta'=\beta$), as we do here,
the reweighting factor $W$ can be written as
\begin{equation}
   W=\frac{1}{\det(A)}\,, \quad 
   A=\prod_{i=1}^{\nf}\frac{D+m_i+i\gamma_5\mu_i}{D+m_i'+i\gamma_5\mu_i'}\label{e:rew-det}\,.
\end{equation} 

The next section specifies the stochastic estimator used for the determinant in
\eqref{e:rew-det}. In section \ref{s:factors} follows a summary of the properties
of three important cases of mass reweighting.
From these properties an optimal strategy for the tuning
of the strange quark mass is derived in section \ref{s:strange-rew}. Section
\ref{s:code} briefly describes a publicly available implementation of this strategy
and in section \ref{s:test-cost} numerical tests and cost estimates are presented.

\section{Fluctuations}
\label{s:fluc-corr}
The numerical estimation of reweighting factors of the form \eqref{e:rew-det}
uses the Monte-Carlo evaluation of an integral representation of the determinant
of a complex matrix \cite{Finkenrath:2013soa,Leder:2014ota}
\begin{equation}\label{e:estimator}
   W_{\Neta}(A) = \frac{1}{\Neta}\sum_{k=1}^{\Neta}
                  e^{-{\eta^{(k)}}^\dagger (A-I) \eta^{(k)}}\,,
\end{equation} 
with $\Neta$ Gaussian distributed random vectors $\eta^{(i)}$ and $W_{\Neta}\to W$
for $\Neta\to\infty$ if $\lambda(A+A^\dagger)>0$ \cite{Finkenrath:2013soa,Leder:2014ota}.
Let us assume $A$ can be written as $A=I+\epsilon B$ with $\epsilon||B||\ll 1$. Then
the stochastic error of this estimator may be expanded
\begin{equation}
   \delta_\eta^2(A)=\var_\eta(W_{\Neta})/(\Neta|W|^2)=\frac{1}{\Neta}[\epsilon^2 \Tr(BB^\dagger) + \rO(\epsilon^3)]\,.
\end{equation}
A similar expansion is obtained for the fluctuations of the reweighting factor
in the expectation value \eqref{e:exp-value} 
\begin{equation}
   \sigma^2_U = \var_U(W)/\ev{W}^2 = \epsilon^2 \var_U(\Tr(B)) + \rO(\epsilon^3)\,.
\end{equation} 

It has been noticed that the stochastic estimator \eqref{e:estimator}
may be plagued by large fluctuations and/or long tails in the distribution.
The reason is that its variance may not be defined even if
the mean is, i.e., the variance is only defined for $\lambda(A+A^\dagger)>1$
\cite{Hasenfratz:2002pt}. There have been several attempts to control the
variance \cite{Hasenbusch:2001ne,Hasenfratz:2002ym,Hasenfratz:2008fg}. Here we
use a factorization of the determinant in $N$ factors with
$\epsilon'\propto\epsilon/N$ \cite{Finkenrath:2013soa,Leder:2014ota}. Since the
factorization has been described thoroughly elsewhere and for better readability
we skip the details here and stick to the formulas with $N=1$. The generalization
is straightforward.

\section{Mass reweighting factors}
\label{s:factors}
\subsection{One flavor}
The simplest case of mass reweighting is the change of the mass of one quark flavor
of mass $m_s$ to mass $m_s+\Dm$. Using the shorthand $D_m=D+m$, the corresponding
reweighting factor is ($\mu_s=0$)
\begin{equation}
   W_\f(m_s,\Dm) = \frac{\det(D_{m_s+\Dm})}{\det(D_{m_s})} = \frac{1}{\det(A_\f)}\,,\quad
      A_\f=I-\Dm D_{m_s+\Dm}^{-1}\,,
\end{equation}                      
and the stochastic error and the fluctuations scale (asymptotically) with $\Dm^2$.

\subsection{Two flavors}
\label{s:2f}
Now we consider the simultaneous change of the mass of two quark flavors $s$ and $l$.
Without loss of generality we assume $m_l\le m_s$ and a change
$m_l\rightarrow m_l-\gamma\Dm$ while $m_s\rightarrow m_s+\Dm$.
The corresponding reweighting factor is ($\mu_s=\mu_l=0$)
\begin{align} \label{e:2f}
   W_\ff^{(\gamma)}(m_l,m_s,\Dm) &= \frac{\det(D_{m_l-\gamma\Dm}D_{m_s+\Dm})}{\det(D_{m_l} D_{m_s})}
            = \frac{1}{\det(A_\ff)}\,,\quad 
      A_\ff =I+\Dm \frac{\gamma\Dm+\gamma D_{m_s}-D_{m_l}}{D_{m_s+\Dm} D_{m_l-\gamma\Dm}} \,,
\end{align}                      
and the stochastic error and the fluctuations scale (asymptotically) with $\Dm^2$.                      
However, if $m_l=m_s=m$ and $\gamma=1$ one finds $A_\ff=I+\Dm^2 (D_m^2-\Dm^2)^{-1}$
and a scaling $\propto \Dm^4$ (see the isospin reweighting in \cite{Leder:2014ota,lat14:jacob}).
In this case the noise and the fluctuations of decreasing the mass of one quark are
compensated by increasing the mass of the second one. For $\gamma=0$ the reweighting
factor reduces to the one flavor case: $W_\ff^{(0)}(m_l,m_s,\Dm)=W_\f(m_s,\Dm)$.
 Therefore, in the general case $m_l< m_s$,
 an optimal $0\le\gamma^*\le 1$ may be found that minimizes the fluctuations of
$W_\ff^{(\gamma)}$. With light quarks around the strange quark mass an optimal
value of $\gamma^*\approx 0.82$ was found \cite{Finkenrath:2013soa}.

A generalization of eq.~\eqref{e:2f} for finite $\mu_s$ and/or $\mu_l$ is straightforward.
Note that, in general, in this case the reweighing factor is complex.

\subsection{Twisted mass reweighting}
\label{s:tm}
Twisted mass reweighting was introduced \cite{Luscher:2008tw} and implemented
\cite{Luscher:2012av} in order to stabilize the HMC at small quark masses. In
the context of mass reweighting it can be used to ensure the convergence of the
stochastic estimator \eqref{e:estimator} \cite{Finkenrath:2013soa}. For a doublet
of quarks of mass $m_1=m_2=m$ and $\mu_1=\mu=-\mu_2$ the corresponding reweighting factor is
\begin{equation}
   W_\tm(m,\mu,\mu') = \frac{\det(D_{m}D_{m}^\dagger+\mu'^2)}{\det(D_{m}D_{m}^\dagger+\mu^2)} = \frac{1}{\det(A_\tm)}\,, \quad
           A_\tm=I+ \frac{\mu^2-\mu'^2} {D_{m}D_{m}^\dagger+\mu'^2}\,,
\end{equation}                      
where we used $(D_{m}+i\gamma_5\mu)^\dagger=\gamma_5 (D_{m}-i\gamma_5\mu) \gamma_5$.
The stochastic error and the fluctuations scale (asymptotically) with $(\mu^2-\mu'^2)^2$.

\section{Optimal strange quark mass reweighting}
\label{s:strange-rew}
\begin{table}[t]
\centering
\begin{tabular}{llllll}
\toprule
       & $(m,\mu)$ & (i)                      & (ii) & (iii) & (iv) \\
\midrule
up & $(m_l,0)$   & $\to (m_l,\mu)$  & $\to (m_l-\gamma\Dm,\mu)$ &  &  $\to (m_l-\gamma\Dm,0)$ \\
down & $(m_l,0)$ & $\to (m_l,-\mu)$  &  & $\to (m_l-\gamma\Dm,-\mu)$  & $\to (m_l-\gamma\Dm,0)$ \\
strange & $(m_s,0)$  &  & $\to (m_s+\Dm,0)$  &  $\to (m_s+2\Dm,0)$  &    \\
\midrule
type     & & tm (sec. \ref{s:tm}) & 2f (sec. \ref{s:2f}) & 2f  (sec. \ref{s:2f}) & tm (sec. \ref{s:tm})\\
\bottomrule
\end{tabular}
\caption{Optimal reweighting strategy for strange quark mass reweighting.}
\label{t:strategy}
\end{table}
As pointed out in the introduction, in a $\nf=2+1$ simulation with up, down, strange
quark masses $\{m_l,\, m_l,\, m_s\}$ it might be necessary
to tune the strange quark mass of an ensemble in order to improve the approach
to the physical point. We have seen in section \ref{s:factors} that it is
advantageous to combine the change of the mass of one quark with an opposite change
of another one. Furthermore, if light quarks are reweighted, a finite twisted mass serves
as a safeguard against zero crossings of small eigenvalues. Taken together, these
lessons lead us to propose the following reweighting strategy:
\begin{enumerate}
 \item[(i)] the light quarks are reweighted to a finite $\mu$,
 \item[(ii)] the up and the strange quark are reweighted together in opposite directions,
 \item[(iii)] the down and the strange quark are reweighted together by the same amount,
 \item[(iv)] the light quarks are reweighted to zero twisted mass.
\end{enumerate}
The four steps are summarized in Table \ref{t:strategy}. In the stochastic estimation
the twisted mass reweighting in step (i) and (iv) can be combined so that the
total reweighting factor $W_s$ can be written as a product of three factors
\begin{equation}\label{e:Ws}
 W_s = \Wii\, \Wiii\, \Wiiv\quad  \leftarrow \quad
      (\Wii)_{\Neta}\, (\Wiii)_{\Neta}\, (\Wiiv)_{\Neta}\,, 
\end{equation}
and each of these factors is estimated according to \eqref{e:estimator}.
The total change in the masses is
$$ \{m_l,\, m_l,\, m_s\} \to \{m_l-\gamma\Dm,\, m_l-\gamma\Dm,\, m_s+2\Dm\}\,. $$

If the ensemble is generated at finite $\mu$ in the light quark sector, as for
example in \cite{Bruno:2014jqa}, step (i) can be omitted. Depending on which kind
of twisted mass reweighting is used in the production of the ensemble, the last factor
is then given by
\begin{align}
   \Wiiv & = \begin{cases}
            W_\tm(m_l,0,\mu)\, W_\tm(m_l-\gamma\Dm,\mu,0) & \text{none}\\
            W_\tm(m_l-\gamma\Dm,\mu,0) & \text{type I in \cite{Luscher:2012av}}\;.\\
            W_\tm(m_l,\mu,\sqrt{2}\mu)\, W_\tm(m_l-\gamma\Dm,\mu,0) & \text{type II in \cite{Luscher:2012av}}
           \end{cases}
\end{align}

The reweighting factors for step (ii) and (iii) are explicitly given by
\begin{align}
   \Wii & = W_\ff^{(\gamma)}(m_l,m_s,\Dm)\,, \quad \text{with $\mu_l=\mu$} \\
   \Wiii & = W_\ff^{(\gamma)}(m_l,m_s+\Dm,\Dm)\,, \quad \text{with $\mu_l=-\mu$}\,,
\end{align} 
and $\gamma\approx\gamma^*$. As mentioned before for $\mu_l\neq 0$ the reweighting
factor $W_\ff^{(\gamma)}$ is complex. But since the product $\Wii \Wiii$ is
manifestly real, the phases of the two factors cancel.

\section{Public code for mass reweighting}
\label{s:code}
The numerical results presented in the next section have been obtained with
the mrw-package \cite{Leder:mrw}, an extended
version of the openQCD package \cite{Luscher:openQCD}. Keeping the structure of
openQCD the extension is implemented as a module (\texttt{mrw}). It provides
a main program that reads in openQCD style input files, documentation
and sample input files.
The mrw-package is publicly available at {\small\url{https://github.com/bjoern-leder/mrw}}.
In detail it adds the following features to \mbox{openQCD}:
\begin{itemize}
   \item one flavor mass and twisted-mass reweighting \cite{Finkenrath:2013soa,Leder:2014ota}
   \item interpolation (factorization) for twisted mass reweighting type I and II 
   \item factorization with non-equidistant interpolations \cite{Finkenrath:2013soa,Leder:2014ota}
   \item isospin mass reweighting \cite{Finkenrath:2013soa,Leder:2014ota}
   \item strange quark mass reweighting of section \ref{s:strange-rew}
   \item several check routines for all parts of the new module
\end{itemize}

\section{Numerical test and cost}
\label{s:test-cost}
\begin{table}[t]
\centering
\begin{tabular}{cccccl}
\toprule
ID in \cite{Bruno:2014jqa} & $\mpi$ [MeV] & $\mK$ [MeV] & $\Delta\mb_s$  [MeV] & $\gamma$ & $\mu$\\ 
\midrule
 -- & 330 (380) & 450 (430) &  -12 & 0.80 & 0.0\\
B105 & 280 (200) & 460 (480) &  \phantom{-}12 & 0.80 & 0.001\\
\bottomrule
\end{tabular}
\caption{Ensembles used in the numerical test. In parentheses are the meson masses 
after reweighting. $\Delta\mb_s$ is the difference of the renormalized strange 
quark masses before and after the reweighting. The first ensemble was part of an exploratory
study and is not listed in \cite{Bruno:2014jqa}.}
\label{t:ensembles}
\end{table}

The optimal strange quark mass reweighting proposed in section \ref{s:strange-rew}
has been tested on two $\nf=2+1$ ensembles from the effort described in
\cite{Bruno:2014jqa}, see Table \ref{t:ensembles}. While the first one is off the
quark mass trajectory described in section 2.3 of \cite{Bruno:2014jqa}, the second
one sits on it.\footnote{This trajectory is defined by $\sum_i m_i = \text{const}$.
Note that for $\gamma\neq 1$ the optimal strange quark
reweighting takes the ensemble off, whereas for $\gamma=1$ it would move
the ensemble along this trajectory.} The lattice spacing is approximated to
$a=0.086\fm$ \cite{Bruno:2014jqa} and the lattice size is $64\times 32^3$.

\begin{figure}
\centering
\includegraphics[width=0.6\textwidth]{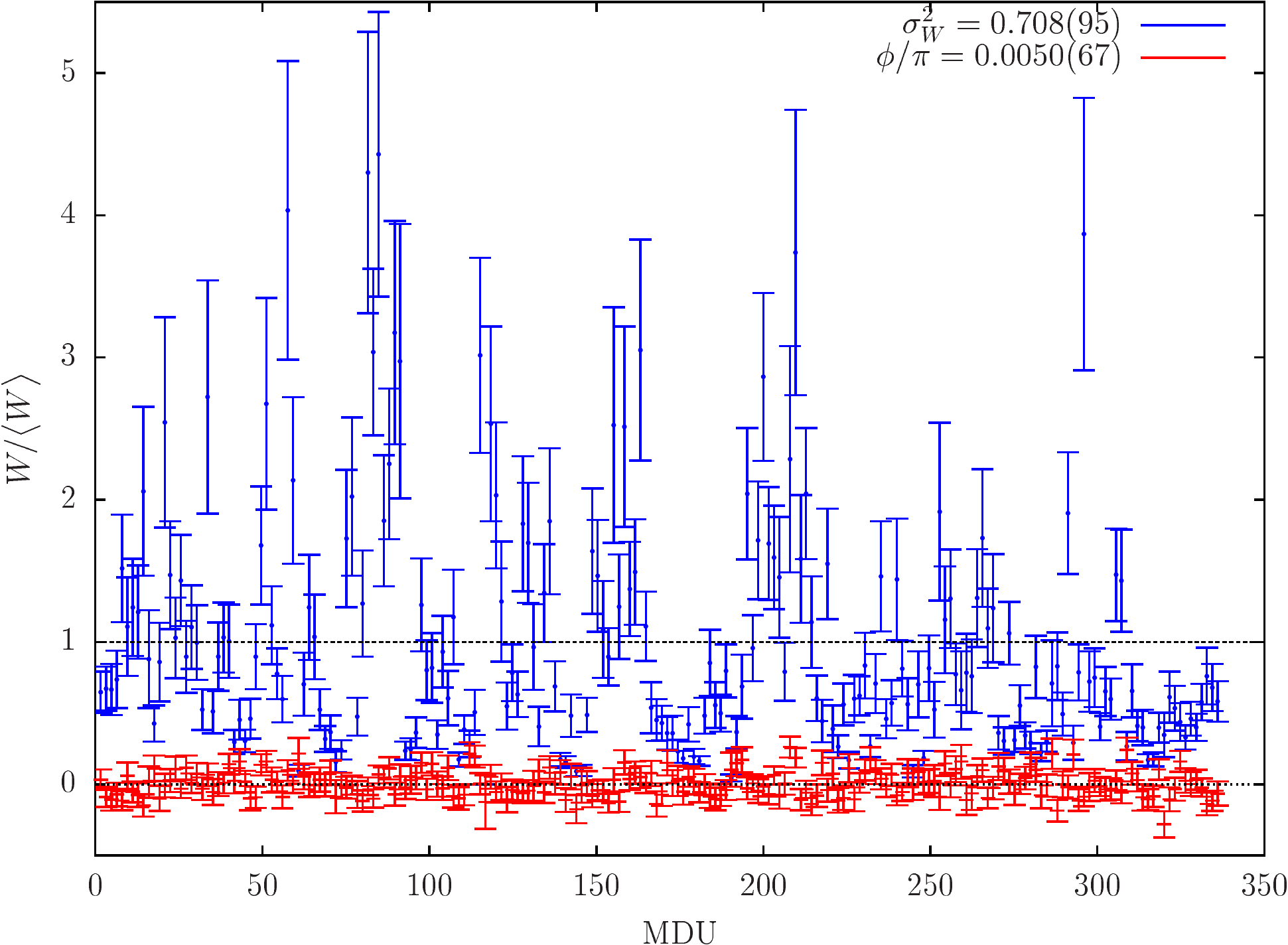}
\caption{Strange quark reweighting factor $W_s$ for ensemble B105.}
\label{f:Ws}
\end{figure} 

In Figure \ref{f:Ws} the reweighting factor $W_s$ (eq. \eqref{e:Ws}) is plotted
for the ensemble B105. The phase is also plotted and is compatible with zero
as expected. The fluctuations of the reweighting are sizable, but the change in
the meson masses is large: 5\% (30\%) for the kaon (pion). By using a smaller
value for $\gamma$ this change can be made more balanced. Assuming a scaling
of the fluctuations $\propto \Dm^2 V^q/\mb_l$ the feasibility of such a
reweighting at different parameters can be projected using the following formula
$$ \sigma^2_{W_s} = 0.71 \, \left(\frac{\Delta\mb_s}{12\,\mev}\right)^2 \,
            \left(\frac{240\,\mev}{\overline{\mpi}}\right)^2 \, 
            \frac{V^q}{(3.3\,\fm)^{4q}}\,, \quad q=1/4 \ldots 3/4\,.$$
where $\Delta\mb_s$ is the difference of the renormalized strange 
quark masses and $\overline{\mpi}$ is the mean of the 
pion masses before and after the reweighting, and $V$ is
the physical volume \cite{lat14:jacob}. The uncertainty in the volume scaling is
currently under investigation.

The cost for estimating the reweighting factors $\Wii$, $\Wiii$, $\Wiiv$ is roughly
independent of the ensemble parameters. It is fixed by demanding the stochastic
noise to be much smaller than the fluctuations \cite{Finkenrath:2013soa}. The error
bars in Figure \ref{f:Ws} indicate the stochastic error for each configuration. 
The total number of Gaussian noise vectors was 48 for $\Wii$ and $\Wiii$, and  24
for $\Wiiv$. Since the stochastic error is large in Figure \ref{f:Ws} these numbers
have to be increased in an application of the method.

The numerical cost per configuration (core hours) is $\sim 30\%$ of one trajectory
of length $2$ MDU. But since the autocorrelation times of observables $\tau_\obs$
are typically much larger than this, reweighting is only needed every
$n=\tau_\obs/2\gg 1$ trajectories, effectively cutting the cost to $\sim (30/n)\%$.

\section{Outlook}
In the first numerical test presented in the last section a value for $\gamma^*$
is used that was determined in another setup and at different quark masses. A dedicated
tuning needs only a small number of configurations and has the potential to lower
the fluctuations significantly. Furthermore the cost per configuration can be
reduced by combining the reweighting factors $\Wii$, $\Wiii$ into one estimator.
Finally, the mrw-package will be upgraded to openQCD-1.4, including support for
periodic boundary conditions.

\paragraph{Acknowledgements:}  This work was funded by the Deutsche
Forschungsgemeinschaft (DFG) in form of Transregional
Collaborative Research Centre 55 (SFB/TRR55).

\bibliographystyle{JHEP}
\bibliography{ref}
%

\end{document}